\documentclass{ws-ijmpa}

\begin{document}

\markboth{Shu Luo and Zhi-zhong Xing} {The Minimal Type-II Seesaw
Model and Flavor-dependent Leptogenesis}

%
\catchline{}{}{}{}{}
%

\title{The Minimal Type-II Seesaw Model and Flavor-dependent Leptogenesis}

\author{Shu Luo$^*$ and Zhi-zhong Xing}

\address{Institute of High Energy Physics, Chinese Academy of
Sciences, Beijing 100049, China \\$^*$E-mail:
luoshu@mail.ihep.ac.cn}

\maketitle

\begin{history}
\received{\today}
\end{history}

\begin{abstract}
Current experimental data allow the zero value for one neutrino
mass, either $m^{}_1 =0$ or $m^{}_3 =0$. This observation implies
that a realistic neutrino mass texture can be established by
starting from the limit (a) $m^{}_1 = m^{}_2 =0$ and $m^{}_3 \neq 0$
or (b) $m^{}_1 = m^{}_2 \neq 0$ and $m^{}_3 =0$. In both cases, we
may introduce a particular perturbation which ensures the resultant
neutrino mixing matrix to be the tri-bimaximal mixing pattern or its
viable variations. We find that it is natural to incorporate this
kind of neutrino mass matrix in the minimal Type-II seesaw model
with only one heavy right-handed Majorana neutrino $N$. We show that
it is possible to account for the cosmological baryon number
asymmetry in the $m^{}_3 =0$ case via thermal leptogenesis, in which
the CP-violating asymmetry of $N$ decays is attributed to the
electron flavor.

\end{abstract}

\ccode{PACS numbers: 11.30.Fs, 14.60.Pq, 14.60.St}

\section{Introduction}

Recent neutrino oscillation experiments have provided us with
convincing evidence that neutrinos are slightly massive and lepton
flavors are not conserved. A global analysis of current neutrino
oscillation data yields $\Delta m^2_{21} = (7.2 \cdots 8.9) \times
10^{-5} ~{\rm eV}^2$ and $\Delta m^2_{32} = \pm (2.1 \cdots 3.1)
\times 10^{-3} ~{\rm eV}^2$ for three neutrino masses as well as
$30^\circ < \theta_{12} < 38^\circ$, $36^\circ < \theta_{23} <
54^\circ$ and $\theta_{13} < 10^\circ$ for three mixing angles at
the $99\%$ confidence level.\cite{Vissani} Three CP-violating phases
remain entirely unconstrained. Among many models or ans$\ddot{a}$tze
proposed to explain the smallness of neutrino masses, the seesaw
mechanism\cite{SS} seems to be most elegant and natural. Associated
with the seesaw idea, the leptogenesis mechanism\cite{FY} turns out
to be an elegant and natural way to interpret the cosmological
baryon number asymmetry which is characterized by $\eta_{B}^{} =
(6.1 \pm 0.2) \times 10^{-10}$.\cite{WMAP}

We find that at least two lessons can be learnt from current
experimental data. First, the lightest neutrino is allowed to be
massless; i.e., either $m^{}_1 =0$ (normal hierarchy) or $m^{}_3 =0$
(inverted hierarchy) has not been excluded. In both cases, the
non-vanishing neutrino masses can be determined in terms of $\Delta
m^2_{21}$ and $|\Delta m^2_{32}|$: (a) if $m^{}_1 = 0$, we have
$m^{}_2 = \sqrt{\Delta m^2_{21}} \approx 8.94 \times 10^{-3} {\rm
eV}$, $m^{}_3  = \sqrt{|\Delta m^2_{32}| + \Delta m^2_{21}} \approx
5.08 \times 10^{-2} {\rm eV}$; or (b) if $m^{}_3 = 0$ we can obtain
$m^{}_1  = \sqrt{|\Delta m^2_{32}| - \Delta m^2_{21}} \approx 4.92
\times 10^{-2} {\rm eV}$ and $m^{}_2 = \sqrt{|\Delta m^2_{32}|}
\approx 5.00 \times 10^{-2} {\rm eV}$. Second, a special neutrino
mixing pattern, the so-called tri-bimaximal mixing,\cite{HPS}
\begin{equation}
V = \left ( \begin{matrix} ~ 2/\sqrt{6} & ~ 1/\sqrt{3} & 0 \cr
-1/\sqrt{6} & ~ 1/\sqrt{3} & ~ 1/\sqrt{2} ~\cr ~ 1/\sqrt{6} &
-1/\sqrt{3} & ~ 1/\sqrt{2} ~\cr \end{matrix} \right ) \; ,
\end{equation}
is particularly favored. It yields $\theta_{12} \approx 35.3^\circ$,
$\theta_{23} = 45^\circ$ and $\theta_{13} = \rho = \sigma =0^\circ$.
As a consequence of $\theta_{13} = 0^\circ$, the Dirac phase
$\delta$ is not well defined. In this talk, we shall first
reconstruct the simplest neutrino mass texture for both mass
hierarchies by combining the above two lessons and then discuss its
seesaw realization and flavor-dependent leptogenesis.

\section{Deviations from Tri-bimaximal neutrino mixing}

Let us work in the basis where the charged-lepton mass matrix
$M^{}_l$ is diagonal. If $V$ is of the tri-bimaximal mixing
pattern as given in Eq. (1), it can be decomposed into a product
of two Euler rotation matrices: $V = O^{}_{23} O^{}_{12}$, where
\begin{eqnarray}
O^{}_{12} = \left ( \begin{matrix} ~ \sqrt{2}/\sqrt{2+x^2} & ~~
x/\sqrt{2+x^2} &~ 0 ~\cr ~ -x/\sqrt{2+x^2} & ~ \sqrt{2}/\sqrt{2+x^2}
&~ 0 ~\cr ~~ 0 & ~ 0 & ~ 1 ~\cr \end{matrix} \right ) \; , ~~~
O^{}_{23} = \left (
\begin{matrix} ~ 1 & ~ 0 & ~ 0 ~\cr ~ 0 & ~ 1/\sqrt{2} & ~ 1/\sqrt{2} ~ \cr ~ 0 &
-1/\sqrt{2} & ~ 1/\sqrt{2} ~ \cr \end{matrix} \right ) \; ,
\end{eqnarray}
with $x=1$. Allowing for small deviations of $x$ from $1$ (or
equivalently deviations of $\theta_{12}^{}$ from $35.3^\circ$), we
are then left with some variations of the tri-bimaximal neutrino
mixing pattern which can fit current neutrino oscillation data if
$0.82 \lesssim x \lesssim 1.10$.

Our strategy of reconstructing the neutrino mass matrix $M^{}_\nu$
is three-fold: (1) we take a proper symmetry limit of $M^{}_\nu$,
denoted as $M^{(0)}_\nu$. $M^{(0)}_\nu$ can be diagonalized by the
transformation $O^{}_{23}$ and the eigenvalues of $M^{(0)}_\nu$
satisfy (a) $m^{}_1 = m^{}_2 =0$ and $m^{}_3 \neq 0$ or (b) $m^{}_1
= m^{}_2 \neq 0$ and $m^{}_3 =0$; (2) we introduce a particular
perturbation to $M^{(0)}_\nu$, denoted as $\Delta M^{}_\nu$, which
can be diagonalized by the transformation $O^{}_{23} O^{}_{12}$; (3)
we require that $M^{}_\nu = M^{(0)}_\nu + \Delta M^{}_\nu$ should
also be diagonalized by the transformation $O^{}_{23} O^{}_{12}$,
and either $m^{}_1 =0$ or $m^{}_3 =0$ is guaranteed.

In the $m^{}_1 =0$ case, the neutrino mass matrix can be written as
\begin{equation}
M^{}_\nu \; = \; M^{(0)}_\nu + \Delta M^{}_\nu \; =\; c \left [
\left ( \begin{matrix} ~ 0 & ~ 0 & ~ 0 ~ \cr ~ 0 & ~ 1 & ~ 1 ~ \cr ~
0 & ~ 1 & ~ 1 ~ \cr
\end{matrix} \right ) + \varepsilon
\left ( \begin{matrix} ~ x^2 & ~ x & -x ~\cr ~ x & ~ 1
& -1 ~\cr -x & -1 & ~1 ~\cr \end{matrix} \right ) \right ] \; .
\end{equation}
Three neutrino masses turn out to be $m^{}_1 =0$, $m^{}_2 = \left
(2+x^2 \right ) c \varepsilon$ and $m^{}_3 = 2c$.

In the $m^{}_3 =0$ case, we have
\begin{equation}
M^{}_\nu \; = \; M^{(0)}_\nu + \Delta M^{}_\nu \; =\; c \left [
\left ( \begin{matrix} ~ 2 & ~ 0 & ~ 0 ~ \cr ~ 0 & ~ 1 & -1 ~ \cr ~
0 & -1 & ~ 1 ~ \cr
\end{matrix} \right ) + \varepsilon \left (
\begin{matrix} ~ x^2 & ~ x & -x ~ \cr ~ x & ~ 1
& -1 ~ \cr -x & -1 & ~ 1 ~ \cr \end{matrix} \right ) \right ] \; ,
\end{equation}
and three neutrino masses are $m^{}_1 = 2c$, $m^{}_2 = \left [2 +
\left ( 2+x^2 \right ) \varepsilon \right ] c$ and $m^{}_3 = 0$.

\section{The minimal Type-II seesaw and $e$-leptogenesis}

Now let us consider how to derive the neutrino mass matrix
$M^{}_\nu$ in Eq. (3) or Eq. (4) from a specific seesaw model.
Taking account of the fact that $M^{}_\nu$ is composed of two mass
matrices $M^{(0)}_\nu$ and $\Delta M^{}_\nu$, we find that it is
quite natural to incorporate $M^{}_\nu$ in the minimal Type-II
seesaw model with only one heavy right-handed Majorana neutrino
$N$.\cite{Gu} Comparing the well-known Type-II seesaw
formula\cite{SS2} $M^{}_\nu \simeq M^{}_L - M^{}_D M^{-1}_R M^T_D$
with $M^{}_\nu = M^{(0)}_\nu + \Delta M^{}_\nu$, we arrive at
\begin{equation}
M^{(0)}_\nu \; = \; M^{}_L \; , ~~~~ \Delta M^{}_\nu \; = \; -M^{}_D
M^{-1}_R M^T_D \; .
\end{equation}
The texture of $\Delta M^{}_\nu$ can be derived from Eq. (5) with a
unique form of $M^{}_D$, $M^{}_D = i\sqrt{c\varepsilon M} \left (
\begin{matrix} x, & 1, & -1 \cr
\end{matrix} \right )^{T}$,
together with $M^{}_R =M$ which is just the mass of the single
right-handed Majorana neutrino. We remark that such a seesaw
realization of the texture of $M^{}_\nu$ does not involve any
fine-tuning or cancellation.

We proceed to consider the leptogenesis in this minimal Type-II
seesaw scenario. We allow $x$ to be complex in $M^{}_D$ and its
imaginary part is just responsible for CP violation in the model.
This assumption surely modifies the pattern of the mixing matrix $V$
given in Section 2, but it maintains $\theta_{13}^{} = 0$ and
$\theta_{23}^{} = 45^{\circ}$.\cite{07043153} In the $m^{}_1 =0$
case, $M^\dagger_D M^{}_L M^{}_D =0$ holds, implying the absence of
CP violation in the decays of $N$. Hence we shall focus our interest
on the $m^{}_3 =0$ case in the following.

In the minimal Type-II seesaw model, the CP-violating asymmetry
between $N\rightarrow l^{}_\alpha + H^{\rm c}$ and its CP-conjugate
process $N\rightarrow l^{\rm c}_\alpha + H$ arises from the
interference between the tree-level amplitude and the
$\Delta^{}_L$-induced one-loop vertex correction, where
$\Delta^{}_L$ denotes the $SU(2)^{}_L$ Higgs triplet. Since our
numerical results indicate that the most favored range of $M$ is
$10^9 ~{\rm GeV} \sim 10^{11} ~{\rm GeV}$, we have to take account
of the flavor effects on leptogenesis.\cite{Flavor} For each lepton
flavor $\alpha$, the corresponding CP-violating asymmetry is
approximately given by\cite{Antusch}
\begin{eqnarray}
\epsilon^{}_{\alpha} \simeq \frac{3 M}{16 \pi v^2} \cdot
\frac{\displaystyle \sum^{}_{\alpha, \beta} {\rm Im} \left [
\left(M_{D}^{*} \right)_{\alpha 1}^{} \left(M_{D}^{*} \right)_{\beta
1}^{} \left( M_{L}^{} \right)_{\alpha\beta}^{} \right ]}{\left(
M_{D}^{\dagger} M_{D}^{} \right)_{11}} \; ,
\end{eqnarray}
where $v \equiv \langle H\rangle \simeq 174$ GeV. In view of $M^{}_L
= M^{(0)}_\nu$ given in Eq. (4) together with $M^{}_D =
i\sqrt{c\varepsilon M} \left ( \begin{matrix} x, & 1, & -1 \cr
\end{matrix} \right )^{T}$, we explicitly obtain
\begin{equation}
\epsilon_{e}^{} \; = \; - \frac{3 M c}{8 \pi v^2} \cdot \frac{{\rm
Im} \left [ \left(x_{}^{*} \right)^2 \right ]}{2 + |x|^2} \; ,
~~~~~~~~ \epsilon_{\mu}^{} \; = \; \epsilon_{\tau}^{} \; = \; 0 \; .
\end{equation}
The overall CP-violating asymmetry turns out to be $\epsilon =
\epsilon^{}_e + \epsilon^{}_\mu + \epsilon^{}_\tau = \epsilon^{}_e$.
This result implies that only the $e$-flavor contributes to
leptogenesis in our model.

The CP-violating asymmetry $\epsilon = \epsilon^{}_e$ can partially
be converted into a net baryon number asymmetry:\cite{Flavor} $
\eta_B^{} \simeq -0.96 \times 10^{-2} \sum^{}_\alpha
\epsilon^{}_\alpha \kappa^{}_\alpha = -0.96 \times 10^{-2}
\epsilon^{}_e \kappa^{}_e$, where the efficiency factors
$\kappa^{}_\alpha$ measure the flavor-dependent washout
effects.\cite{BBP} In the model under discussion, we have
$\kappa^{}_e \simeq (0.467\cdots 0.64)$. The feasibility of
producing $\eta^{}_B \sim 6 \times 10^{-10}$ has been confirmed by
our numerical analysis.\cite{07043153}

\section{Summary}

In summary, we have proposed a new category of neutrino mass
ans${\rm\ddot{a}}$tze by starting from a combination of two
phenomenological observations: (1) the lightest neutrino mass might
be zero or vanishingly small, and (2) the neutrino mixing matrix
might be the tri-bimaximal mixing pattern or a pattern close to it.
We have incorporated the texture of $M^{}_\nu$ in the minimal
Type-II seesaw model and demonstrated that our model can
simultaneously interpret current neutrino oscillation data and the
cosmological baryon number asymmetry via thermal leptogenesis, in
which only the electron flavor plays a role in the lepton-to-baryon
conversion.

Finally let us remark that both the neutrino mass spectrum and the
flavor mixing angles are well fixed and are stable against radiative
corrections\cite{RGE} in the proposed model. It is therefore easy to
test them in the near future, when more accurate experimental data
are available.

We would like to thank A. H. Chan and H. Fritzsch for
collaboration\cite{07043153}. This work was supported in part by the
National Natural Science Foundation of China.

\end{document}